\documentclass{amsart}[10pt]

\usepackage{amscd,amssymb,latexsym,url,verbatim,graphicx,color}
\usepackage{tikz,tikz-cd}
\usetikzlibrary{decorations.pathmorphing}

\usepackage{cases,amsmath}

\usepackage{txfonts}

\usepackage{tikz,tikz-cd}

\usepackage[dvips]{epsfig}

\title{ All binomial identities are orderable}

\author{Dmitry N. Kozlov}

\address{Department of Mathematics, University of Bremen, 28334
  Bremen, Federal Republic of Germany}

\email{dfk@math.uni-bremen.de}

\newtheorem{theorem}{Theorem}[section]
\newtheorem{df}[theorem]{Definition}
\newtheorem{thm}[theorem]{Theorem} 
\newtheorem{prop}[theorem]{Proposition}
 \newtheorem{crl}[theorem]{Corollary}

 \newcommand{\nin}{\noindent}
\newcommand{\pr}{\nin{\bf Proof.} }

\newcommand{\ca}{{\mathcal A}}
\newcommand{\cb}{{\mathcal B}}
\newcommand{\cc}{{\mathcal C}}

\newcommand{\cs}{{\mathcal S}}

\newcommand{\cx}{{\mathcal X}}
\newcommand{\cy}{{\mathcal Y}}
\newcommand{\cz}{{\mathcal Z}}

\newcommand{\ra}{\rightarrow}

\newcommand{\gr}{\Gamma_{\ca,\cb}}
\newcommand{\sh}{{\rm Sh}}

\numberwithin{equation}{section}
\numberwithin{figure}{section}
\numberwithin{table}{section}

\begin{document}

\begin{abstract}
The main result of this paper is to show that all binomial identities
are orderable.  This is a natural statement in the combinatorial
theory of finite sets, which can also be applied in distributed
computing to derive new strong bounds on the round complexity of the
weak symmetry breaking task.

Furthermore, we introduce the notion of a fundamental binomial
identity and find an infinite family of values, other than the prime
powers, for which no fundamental binomial identity can exist.
\end{abstract}

\maketitle

\section{Preliminaries}

\nin
For any natural number $n$, we set $[n]:=\{1,\dots,n\}$.

\begin{df}
A~{\bf binomial identity} is any equality
\begin{equation} \label{eq:bin}
\binom{n}{a_1}+\dots+\binom{n}{a_k}=\binom{n}{b_1}+\dots+\binom{n}{b_m},
\end{equation}
where $n$ is a natural number and $0\leq a_1<\dots<a_k\leq n$, $0\leq
b_1<\dots<b_m\leq n$, $a_i\neq b_j$, $\forall i,j$.
\end{df}

Given a~binomial identity~\eqref{eq:bin}, we associate to it the following data:
\begin{itemize} 
\item index sets $A:=\{a_1,\dots,a_k\}$ and $B:=\{b_1,\dots,b_m\}$; 
\item the families of subsets $\ca:=\{S\subseteq[n]\,|\,|S|\in A\}$ and
$\cb:=\{T\subseteq[n]\,|\,|T|\in B\}$. 
\end{itemize}
The binomial identity then simply says that $|\ca|=|\cb|$, i.e., there
exists a~bijection between $\ca$ and $\cb$.

\begin{df}
We say that the binomial identity \eqref{eq:bin} is {\bf orderable} if
there exists a~bijection $\Phi:\ca\ra\cb$, such that for each $S\in\ca$
we either have $S\subseteq\Phi(S)$ or $S\supseteq\Phi(S)$.
\end{df}

One may view $\ca$ and $\cb$ as subsets of the boolean algebra
$\cc^n$, consisting of entire levels indexed by $A$ and $B$. The
binomial identity is then orderable if and only if there is a~perfect
matching between elements of $\ca$ and elements of $\cb$, such that we
are allowed only to match comparable elements. Our main result says
that this can always be done.

\begin{thm}\label{thm:main}
All binomial identities are orderable.
\end{thm}

We note a~special binomial identity $\binom{n}{k}=\binom{n}{n-k}$, for
which Theorem~\ref{thm:main} is well-known and many explicit
bijections have been constructed, e.g., using Catalan factorization
of walks. 
Before we can give our proof of the main theorem, we need to make some
constructions and to recall a few facts.

We start with some graph terminology. Given a~graph $G$, we let $V(G)$
denote its set of vertices, and we let $E(G)$ denote its set of
edges. For a~vertex $v\in V(G)$, we set $N(v):=\{w\in
V(G)\,|\,(v,w)\in E(G)\}$, the set of all vertices {\it adjacent}
to~$v$.  We extend this notation to sets of vertices $S\subseteq V(G)$
by setting $N(S):=\bigcup_{v\in S}N(v)$, so $N(S)$ is the set of all
vertices of $G$ adjacent to {\it some} vertex of~$S$.

A~graph $G$ is called {\it bipartite} if its set of vertices can be
split as a~disjoint union $V(G)=U\cup W$, such that every edge of $G$
has one vertex in $U$ and one vertex in $W$. We shall say that
$G=(U,W)$ is a~{\it bipartite split}; note that it may not be unique
if the graph is not connected. Note that if $S\subseteq U$, then
$N(S)\subseteq W$ and vice versa.

\begin{df}
Assume we are given two disjoint collections of subsets
$\cx,\cy\subseteq 2^{[n]}$. We let $\Gamma_{\cx,\cy}$ denote the
bipartite graph defined as follows:
\begin{itemize}
\item the vertices are the sets in these collections:
  $V(\Gamma_{\cx,\cy})=\cx\cup\cy$;
\item the sets $S\in\cx$ and $T\in\cy$ are connected by an edge if and only
if $S\subset T$ or $T\subset S$.
\end{itemize}
\end{df}

As said above, a~binomial identity is orderable if and only if the
bipartite graph $\gr$ has a~perfect matching. The {\it matching
  theory} is a~rich theory, and the following theorem provides a
standard criterion for the existence of a perfect matching, see e.g.,
\cite{Ca,LP}.

\begin{thm} (Hall's Marriage Theorem).

\nin Assume $G=(A,B)$ is a bipartite graph, such that $|A|=|B|$. The
graph $G$ has a~perfect matching if and only if for every set
$Z\subseteq A$ we have
\begin{equation} \label{eq:marriage}
|N(Z)|\geq|Z|.
\end{equation}
\end{thm}

In addition to the graph terminology, we need some combinatorial
notions related to Boolean algebra. For all $0\leq k\leq n$, we let
$\cc_k^n:=\{S\subseteq[n]\,|\,|S|=k\}$ denote the $k$-th level in the
boolean algebra~$\cc^n$.

\begin{df}
Assume we are given  $\cs\subseteq\cc_a^n$ and $0\leq b\leq n$. 
We set
\[\sh_b(\cs):=\begin{cases}
\{T\in\cc_b^n\,|\,T\subseteq S,\textrm{ for some }S\in\cs\},& \text{ if } b\leq a;\\
\{T\in\cc_b^n\,|\,T\supseteq S,\textrm{ for some }S\in\cs\},& \text{ if } b\geq a.
\end{cases}\] 

Furthermore, for any set $B\subseteq\{0,\dots,n\}$, we set
$\sh_B(\cs):=\bigcup_{b\in B}\sh_b(\cs)$. We call these sets {\bf
  $b$-shadow} and {\bf $B$-shadow} of~$\cs$.
\end{df}

We adopted here the standard terminology from Sperner theory, see
\cite[Chapter 2]{An}, though we do not distinguish between shadows and
shades.  Clearly, in terms of the bipartite graph $\Gamma_{\cx,\cy}$
above, the shadow operation coincides with the adjacency operation
$N(-)$.

\section{The proof of the main theorem}

The crucial fact which we need for our proof is the following result
of Sperner.

\begin{thm} \label{thm:sper} (Local LYM inequality). 

\nin Assume $n$ is an arbitrary natural number and we are given
$\cs\subseteq\cc_a^n$, for some $0\leq a\leq n$. Let
$F:=|\cs|/\binom{n}{a}$ denote the fraction of the chosen $a$-subsets,
then for any $b$ we have
\begin{equation} \label{eq:sh}
|\sh_b(\cs)|\geq  F\cdot\binom{n}{b}.
\end{equation}

Moreover, we have strict inequality in~\eqref{eq:sh}, in case $S$
is a~proper non-empty subset of~$\cc_a^n$.
\end{thm}

Stated colloquially, the local LYM inequality simply says that when
viewed {\it proportionally}, the shadow of the set family $\cs$ is at
least as large as the family $\cs$ itself. This is a~standard result
in Sperner Theory, which can be found e.g., in
\cite[Section~2.1]{An}. One says that the Boolean algebra has the {\it
  normalized matching property}.

For the sake of being self-contained we sketch a simple
double-counting argument proving Theorem~\ref{thm:sper}. Assume for
simplicity that $b\geq a$, the case $a\geq b$ is completely
analogous. Set
$\Lambda:=\{(S,T)\,|\,S\in\cs,\,T\in\cc_b^n,\,S\subseteq T\}$; this is
the set which we want to double-count. On one hand, each $S\in\cs$ is
contained in exactly $\binom{n-a}{b-a}$ subsets of cardinality $b$, so
$|\Lambda|=|\cs|\cdot\binom{n-a}{b-a}$. On the other hand, each set
$T\in\sh_b(\cs)$ contains $\binom{b}{a}$ subsets of cardinality $a$,
though not all of them must be in $\cs$, so
$|\sh_b(\cs)|\cdot\binom{b}{a}\geq|\Lambda|$. Combining these yields
the inequality~\eqref{eq:sh}, since we have the identity
$\binom{n}{b}\binom{b}{a}=\binom{n}{a}\binom{n-a}{b-a}$. The latter is
just the formal way of saying that to choose an $a$-set inside
a~$b$-set, inside a~fixed $n$-set, one can either first pick a~$b$-set
inside that $n$-set, and then an~$a$-set inside the chosen $b$-set, or
first pick an~$a$-set inside the $n$-set, and then complement it to
a~$b$-set, by choosing a~$(b-a)$-set inside the $(n-a)$-set.

Note, that if we get equality in~\eqref{eq:sh}, then we must have
$|\sh_b(\cs)|\cdot\binom{b}{a}=|\Lambda|$. In other words, for every
set $T\in\sh_b(\cs)$, all of its $a$-subsets must be in~$\cs$.  This
means, that if $A\in\cs$, and $A'$ is obtained from $A$ by replacing
a~single element, then $A'\in\cs$ as well. Hence, if
$\cs\neq\emptyset$, then $\cs=\cc_a^n$.

Assume now we are given a set $B\subseteq\{0,\dots,n\}$. Then
\begin{equation} \label{eq:shb}
|\sh_B(\cs)|=\sum_{b\in B}|\sh_b(\cs)|\geq F\cdot\sum_{b\in B}\binom{n}{b}=
F\cdot|\cb|,
\end{equation}
where $\cb:=\{T\,|\,|T|\in B\}$. Again, we get equality
in~\eqref{eq:shb} if and only if $\cs=\emptyset$ or $\cs=\cc_a^n$.


We are now ready to prove our main theorem.

\vskip5pt

\nin {\bf Proof of Theorem~\ref{thm:main}.}  Assume we have a binomial
identity with associated index sets $A$ and $B$, and associated
collections of sets $\ca$ and $\cb$ as described above.  We show that
the bipartite graph $\gr$ has a perfect matching by checking the
condition of the marriage theorem. Let $\cz\subset\ca$,
$\cz\neq\emptyset$, and write $\cz=\cz_1\cup\dots\cup\cz_k$, where
$\cz_i:=\cz\cap\cc_{a_i}^n$.

Set $F_i:=|\cz_i|/\binom{n}{a_i}$, this is the fraction of all
$a_i$-subsets contained in $\cz$. Choose $1\leq r\leq k$ for which
$F_r=\max_i F_i$. If $F_r=1$, then $\sh_B(\cz)=\cb$, so $|\sh_B(\cz)|=
|\cb|=|\ca|>|\cz|$.

Assume now that $F_r<1$, i.e., $\cz_r\neq\cc_{a_r}^n$.  We have
a~chain of equalities and inequalities:
\[|\sh_B(\cz)|\geq |\sh_B(\cz_r)|> F_r\cdot |\cb|= F_r\cdot|\ca|=
F_r\cdot\sum_{i=1}^k\binom{n}{a_i}\geq \sum_{i=1}^k
F_i\cdot\binom{n}{a_i}= \sum_{i=1}^k|\cz_i|=|\cz|,\] where the second
inequality is \eqref{eq:shb}, and all the other steps are
straightforward.  This confirms \eqref{eq:marriage}, hence the perfect
matching exists. 

We remark, that  we actually proved a~slightly stronger condition than
required by the Marriage Theorem. Namely, we have shown that 
\begin{equation}\label{eq:strong}
|\sh_B(Z)|\geq|Z|+1,
\end{equation} 
whenever $Z\neq\emptyset$, $Z\neq\ca$; cf. {\it surplus}
in~\cite{LP}.  \qed

\section{Applications to distributed computing}


In order to keep our presentation compact, this section is not made
self-contained and we shall use terminology and framework
from~\cite{HKR,wsb12}. This section does not contain new mathematical
results, and can be skipped by the reader interested in binomial
identites only. Our application concerns the complexity of solving
Weak Symmetry Breaking (WSB) using Iterated Immediate Snapshot (IIS)
model, see \cite{ACHP,CR2,wsb6,wsb12} for previous work.

One of the central results proved in~\cite{wsb12}, stated that the
existence of certain combinatorial bijections between families of sets
implies the existence of fast IIS protocols solving the WSB task.
Here is the formulation of this result using the language the present
paper.

\begin{thm}\label{thm:c}\cite[Theorem C]{wsb12}. $\,$

\nin
 Assume that for a~certain natural number~$n$ we have a~binomial
identity
\begin{equation} \label{eq:be2}
\binom{n}{0}+\binom{n}{a_1}+\dots+\binom{n}{a_k}=
\binom{n}{1}+\binom{n}{b_1}+\dots+\binom{n}{b_m},
\end{equation}
such that $2<a_1<\dots<a_k<n$, $1<b_1<\dots<b_m<n$, and $a_i\neq b_j$,
for all $i,j$. Set $A:=\{0,a_1,\dots,a_k\}$, $B:=\{1,b_1,\dots,b_m\}$,
$\ca:=\{S\subseteq[n]\,|\,|S|\in A\}$ and
$\cb:=\{T\subseteq[n]\,|\,|T|\in B\}$. Assume furthermore that
there exists a~bijection $\Phi:\ca\ra\cb$, such that 
\begin{itemize}
\item $\Phi(\emptyset)=\{n\}$, 
\item for all $S\in\ca$ we have either $\Phi(S)\subseteq S$ or
  $\Phi(S)\supseteq S$.
\end{itemize}
Then there exists a~$3$-round IIS protocol solving WSB for $n$
processes.
\end{thm}


Clearly, presenting such a~bijection is a~nice combinatorial way to
show the existence of the distributed protocol. Still, as the example
for $n=6t$ in \cite{wsb12} demonstrates, the explicit construction of
this bijection can be a~formidable task.  Fortunately, combining
Theorem~\ref{thm:c} with our main Theorem~\ref{thm:main} we arrive at
the following statement, which allows us to conclude that these
protocols exist based on the numerical evidence alone.


\begin{thm}\label{thm:main2}
Assume $n$ is a natural number, and there exists a~binomial
identity~\eqref{eq:be2}, such that $2<a_1<\dots<a_k<n$,
$1<b_1<\dots<b_m<n$, and $a_i\neq b_j$, for all $i,j$. Then there
exists a~3-round IIS protocol solving WSB for $n$ processes.
\end{thm}
\pr We can construct the desired bijection $\Phi$ as follows.  To
start with, set $\Phi(\emptyset):=\{n\}$. Let $\widetilde\Gamma$ be
obtained from $\Gamma_{\ca,\cb}$ by deleting $\emptyset$ and $\{n\}$.
In the proof of Theorem~\ref{thm:main} we have actually showed that
$\Gamma_{\ca,\cb}$ satisfies the stronger condition~\eqref{eq:strong}.
This means that the graph $\widetilde\Gamma$ satisfies the Marriage
theorem condition and hence has a perfect matching. Thus we get a
bijection $\Phi$ satisfying all the necessary conditions.  \qed

\vskip5pt

The binomial identities of the type~\eqref{eq:be2} exist for all
$n=6t$, where $t$ is an arbitrary natural number:
$\sum_{k=0}^{t-1}\binom{n}{3k}=\sum_{k=0}^{t-1}\binom{n}{3k+1}$, see
also~\cite{wsb12}. This shows that there infinitely many values of $n$
for which such identity exists. On the other hand, when $n$ is a~prime
power $p$, the left hand side of \eqref{eq:be2} is equal to $1$ modulo
$p$, whereas the right hand side is divisible by $p$, so no such
identity can exist.

Here are examples of binomial identities for $n=15$, $20$, and $21$,
satisfying conditions in Theorem~\ref{thm:main2}:
\[\binom{15}{0}+\binom{15}{4}+\binom{15}{6}+\binom{15}{13}=
\binom{15}{1}+\binom{15}{3}+\binom{15}{5}+\binom{15}{10},\] 

\[\binom{20}{0}+\binom{20}{5}+\binom{20}{6}+\binom{20}{7}+\binom{20}{18}=
\binom{20}{1}+\binom{20}{3}+\binom{20}{4}+\binom{20}{8},\]

\[\binom{21}{0}+\binom{21}{4}+\binom{21}{5}+\binom{21}{7}+\binom{21}{14}+\binom{21}{19}=
\binom{21}{1}+\binom{21}{3}+\binom{21}{6}+\binom{21}{8}.\]

\nin The Theorem~\ref{thm:main2} implies then that Weak Symmetry
Breaking can be solved in $3$ rounds for $n=15$, $20$, and~$21$.

\section{Fundamental binomial identities}

We now fix a certain family of binomial identities whose existence
appears to be an interesting but difficult question, which is also
important in various contexts.

\begin{df}\label{df:fund}
Let $n$ be an~arbitrary natural number. A~binomial identity of the
form
\begin{equation} \label{eq:be3}
\binom{n}{0}+\binom{n}{a_1}+\dots+\binom{n}{a_k}=
\binom{n}{b_1}+\dots+\binom{n}{b_m},
\end{equation}
such that $1\leq a_1<\dots<a_k<n$, $1\leq b_1<\dots<b_m<n$, is called
a~{\bf fundamental} binomial identity associated to~$n$.
\end{df}

\nin {\bf Question.} {\it For which values of $n$ does 
a~fundamental identity associated to $n$ exist? }

\vskip5pt

\nin A~fundamental binomial identity certainly does not exist for the
degenerate case $n=1$. Furthermore, when $n=p^\alpha$ is a~prime
power, we see that $p$ divides the right hand side of \eqref{eq:be3},
while it does divide the left hand side.  So again, no identity
exists, and $n$ must be at least $6$.

Clearly, any binomial identity of the type~\eqref{eq:be2} is
fundamental, so from the previous section we know that such an
identity exists for $n=6$. Furthermore, the nonexistane of fundamental
identities will also imply the nonexistance of identities of
type~\eqref{eq:be2}.\footnote{At the time writing we are not aware of
  any $n$ for which a~fundamental binomial identity exists, but an
  identity of the type~\eqref{eq:be2} does not exist.}  The first
value for which no fundamental binomial identity exists is $n=10$. It
is easy to prove the following more general proposition.
\begin{prop}\label{prop:no}
Assume that $n=p^k\cdot q^m$, where $p$ and $q$ are different prime
numbers, $k\geq 1$, and $m\geq 0$. If we have
\begin{equation}\label{eq:pq}
p\geq 2^{q^m},
\end{equation}
then there does not exist any binomial
identity of the type~\eqref{eq:be2}.
\end{prop} 
\pr The case $m=0$ has already been settled, so assume $m\geq 1$. Note
that $p$ divides $\binom{n}{t}$, unless $t=\alpha p^k$. Furthermore,
$\binom{n}{\alpha p^k}\equiv\binom{q^m}{\alpha}\mod p$, for all
$\alpha=0,\dots,q^m$. So a~binomial identity of the
type~\eqref{eq:be2} would imply that we have
\begin{equation}\label{eq:no1}
\epsilon_1\binom{q^m}{1}+\dots+\epsilon_{q^m-1}\binom{q^m}{q^m-1} 
\equiv 1 \mod p,
\end{equation}
for some $\epsilon_1,\dots,\epsilon_{q^m-1}\in\{-1,0,1\}$. On the
other hand, we have
\begin{equation}\label{eq:no2}
\left|\epsilon_1\binom{q^m}{1}+\dots+\epsilon_{q^m-1}\binom{q^m}{q^m-1}\right|
\leq 2^{q^m}-2\leq p-2,
\end{equation}
where the last inequality uses our assumption~\eqref{eq:pq}.
Combining \eqref{eq:no1} with \eqref{eq:no2} we obtain
\[\epsilon_1\binom{q^m}{1}+\dots+\epsilon_{q^m-1}\binom{q^m}{q^m-1}=1,\]
which is impossible, since the left hand side is divisible by $q$.
\qed

As mentioned above, when $m=0$ in Proposition~\ref{prop:no}, we
recover the case when $n$ is a~prime power. When $m=1$, $q=2$, we get
the case $n=2p^k$, for $p\geq 5$, covering the special cases $n=10$,
$14$, $22$, $\dots$. Further special values of $q$ and $m$ will yield
the cases $n=3p^k$, for $p\geq 11$, $n=4p^k$, for $p\geq 17$,
$n=5p^k$, for $p\geq 37$, $n=7p^k$, for $p\geq 131$, etc.

\section{Final remarks}

Curtis Greene, \cite{Gr}, has suggested that the following stronger
version of \cite[Conjecture 11.5]{wsb12} might be true.  Our argument
above yields this more general result as well.

\begin{crl}\label{crl:gr}
Assume that we have an inequality
\[\binom{n}{a_1}+\dots+\binom{n}{a_k}\leq\binom{n}{b_1}+\dots+\binom{n}{b_m}.\]
Set $\Sigma:=\{S\,|\,|S|\in\{a_1,\dots,a_k\}\}$,
$\Lambda:=\{T\,|\,|T|\in\{b_1,\dots,b_m\}\}$.  Then there exists an
injection $\Phi:\Sigma\ra\Lambda$ with $S\subseteq\Phi(S)$ or
$S\supseteq\Phi(S)$, for all $S\in\Sigma$.
\end{crl}
\pr Follows immediately from our argument together with a stronger
version of the Hall's Marriage Theorem, e.g., see~\cite[Theorem 1.3.1]{LP},
since the corresponding bipartite graph here has deficiency~$0$.  
\qed

\vskip5pt

As a final note, we would like to remark that Hall's Marriage Theorem
has many constructive proofs, see, e.g.,\cite{LP}. This means that once we
have an identity~\eqref{eq:be2}, there is a way to {\it construct} the
corresponding perfect matching. That in turn, combined with the direct
path construction in~\cite{wsb12}, yields an explicit 3-round
distributed protocol solving the Weak Symmetry Breaking for $n$
processes.



\vskip10pt

\nin{\bf Acknowledgments.} We thank Curtis Greene for engaged discussions.


\end{document}